# Radiative energy loss in the absorptive QGP: taming the long formation lengths in coherent emission *

M. Bluhm[1,2], P. B. Gossiaux[1,†], T. Gousset[1] and J. Aichelin[1]

[1] SUBATECH, UMR 6457, Université de Nantes, Ecole des Mines de Nantes, IN2P3/CNRS. 4 rue Alfred Kastler, F-44307 Nantes cedex 3, France
[2] CERN, Physics Department, Theory Division, CH-1211 Geneva 23, Switzerland

In an absorptive plasma, damping of radiation mechanisms can influence the bremsstrahlung formation in case of large radiation formation lengths. We study qualitatively the influence of this effect on the gluon bremsstrahlung spectrum off heavy quarks in the quark-gluon plasma. Independent of the heavy-quark mass, the spectrum is found to be strongly suppressed in an intermediate gluon energy region which grows with increasing gluon damping rate and increasing energy of the heavy quark. Thus, just as polarization effects in the plasma render the bremsstrahlung spectra independent of the quark mass in the soft gluon regime, damping effects tend to have a similar impact for larger gluon energies.

## 1. Introduction

Knowing the energy loss mechanisms taking place in the strongly coupled quark-gluon fluid is essential for our understanding of the observed jet quenching and high-$p_T$ hadron suppression as found in ultra-relativistic heavy-ion collisions [1, 2, 3, 4, 5]. It is commonly accepted that the radiative energy loss contribution [6, 7, 8, 9, 10, 11] dominates over the collisional contribution [12, 13, 14]. In the theoretical studies of medium-induced radiative energy loss, however, the influence of a possible damping of gluon bremsstrahlung has barely been taken into account.

Important in the context of discussing radiative energy loss is the notion of a formation time or length of the radiation. This is because a (gluon) bremsstrahlung process from an emitter (parton) is a quantum phenomenon and as such not instantaneous. In case the system is perturbed within the course of this process by the presence of the medium, the bremsstrahlung cross-section will be affected.

---

* Contribution to *Excited QCD 2012*, Peniche, Portugal, May 6-12, 2012.
† presenter

In fact, in a medium a multitude of different effects can potentially disturb the formation of radiation. For instance, if the formation length $l_f$ is large compared to the in-medium scattering mean free path $\lambda$, then emitted quanta will not resolve different scatterings and the assumption of independent emissions will become invalid. Rather, $l_f/\lambda \gg 1$ scatterings will contribute coherently to the radiation process. As a consequence, the radiation spectrum will be reduced compared to the one from independent, successive scatterings. This is known as the LPM-effect [15, 16] in electrodynamics with an analog in QCD, cf. [7].

Likewise, the radiation quanta can Compton-scatter with the medium constituents. This can be described by attributing to the medium a dielectric function different from the one in vacuum. Consequently, quanta emitted within the medium aquire an effective in-medium mass which changes their dispersion relation. This effect, known as the TM-effect [17] in electrodynamics, with an analog in QCD as studied in [18, 19], leads also to a significant reduction of the radiation spectrum. Moreover, if the formation length of the energetic quanta is large compared to, for instance, the mean free path for pair creation in the medium, then this process can influence the radiation formation and therefore the bremsstrahlung spectrum.

In this work, we discuss qualitatively the possible effect of such damping mechanisms like pair creation on the gluon bremsstrahlung spectrum off heavy quarks (HQ). In section 2, the formation length for gluon bremsstrahlung off massive partons is introduced, while in section 3 the impact of gluon damping on the radiation spectrum is investigated. Our results are analyzed and summarized in section 4.

## 2. Formation length of gluon bremsstrahlung

An estimate for the formation length of medium-induced gluon radiation in a strongly interacting medium can be obtained, in analogy with [20], by considering the imbalance in parallel momenta in a bremsstrahlung process which contains a relativistic parton with energy $(1-x)E$ and momentum $\vec{p}$ plus a bremsstrahlung gluon with energy $\omega = xE$ and momentum $\vec{k}$ in the final state and a relativistic single parton intermediate state with the same total energy $E$ and the same total transverse momentum $\vec{p}_\perp + \vec{k}_\perp$. Then, $l_f$ is approximated by the inverse expectation value of that parallel momentum imbalance as a measure for the parallel momentum transfer from the medium to the parton-gluon system.

Considering the special case in which $\vec{p}_\perp = 0$, while the bremsstrahlung gluon accumulates the squared transverse momentum $\langle \vec{k}_\perp^2 \rangle = \hat{q} l_f$ during its formation due to multiple, elastic scatterings, one finds as condition determining $l_f$ for not too small $(1-x)$

$$1 \simeq l_f \left[ \frac{m_g^2}{xE + k_\parallel} + \frac{xm_s^2}{2E(1-x)} \right] + l_f^2 \frac{\hat{q}}{2xE} \left[ \frac{2xE}{xE + k_\parallel} - x \right]. \quad (1)$$

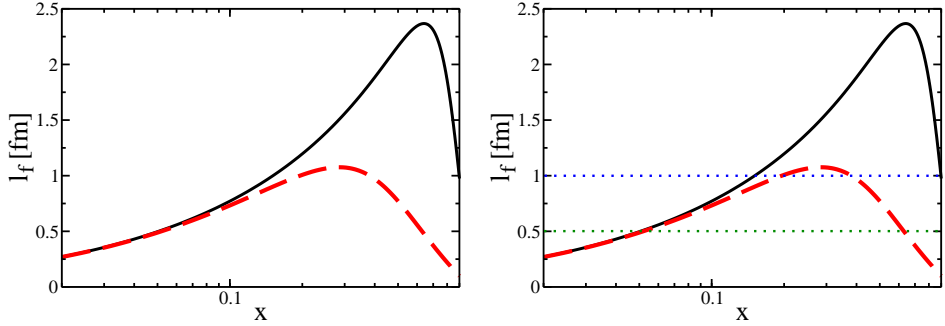

Fig. 1. Left: Formation length $l_f$ of charm quarks (solid curve) and bottom quarks (dashed curve) as a function of gluon energy fraction $x$. The used parameters read $E = 40$ GeV, $m_c = 1.3$ GeV, $m_b = 4.2$ GeV and $m_g = 0.8$ GeV, $\hat{q} = 2$ GeV$^2$/fm. Right: In addition, the absorption length $l_d \simeq 1/\Gamma$ is shown with $\Gamma = 0.2$ GeV (upper dotted curve) and $\Gamma = 0.4$ GeV (lower dotted curve), cf. text for details.

Here, $m_s$ and $m_g$ are the in-medium masses of the parton and of the bremsstrahlung gluon, respectively, while $\hat{q}$ denotes the mean squared transverse momentum per unit length picked up by the gluon. For large $k_\parallel \simeq xE$, equation (1) renders into [21]

$$1 \simeq l_f \left[ \frac{m_g^2}{2xE} + \frac{xm_s^2}{2E(1-x)} \right] + l_f^2 \frac{\hat{q}}{2xE}(1-x). \quad (2)$$

The positive solution for the gluon formation length as a function of $x$ obtained from Eq. (2) agrees with the one obtained from Eq. (1) apart from the region of rather small $x \simeq \sqrt{m_g^2 + \hat{q}l_f}/E$ compared to $E$. In this case, i.e. when $k_\parallel \ll xE$, $m_g^2/(xE + k_\parallel) \to m_g^2/(xE)$ in Eq. (1) and $2xE/(xE + k_\parallel) \to 2$ which, however, merely introduces small changes in the numerical factors compared to Eq. (2).

In Fig. 1, the formation length of gluons radiated off charm and bottom quarks is shown as determined from Eq. (2). For small $x \ll m_g/m_s \ll 1$, the result for $l_f$ is evidently quark mass independent. This is a consequence of the finite in-medium mass of the radiated gluon. Apart from this small-$x$ region, however, $l_f$ exhibits clearly a strong parton mass dependence.

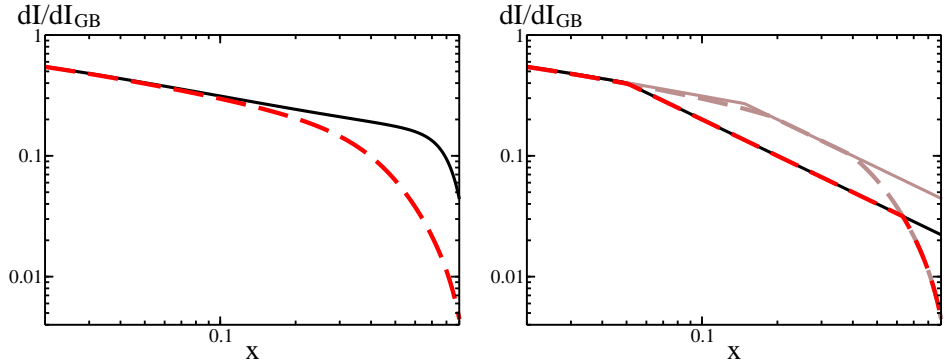

Fig. 2. Left: Differential radiation intensity scaled by the Gunion-Bertsch reference spectrum [24] from a single, high-energy scattering process in the limit of small gluon energy fraction $x \ll m_g/m_s$ shown as a function of $x$ for charm (solid curve) and bottom (dashed curve) quarks. The same parameters as in Fig. 1 are used. Right: Semi-quantitative study of the additional influence of gluon damping phenomena on the scaled radiation intensity of the left panel. Upper (lower) set of curves shows the additional spectrum reduction in case $\Gamma = 0.2$ GeV ($\Gamma = 0.4$ GeV) for charm (solid curves) and bottom (dashed curves) quarks.

Moreover, it is interesting to note that the formation length for a given $x$ increases with parton energy $E$ but decreases with increasing $\hat{q}$.

### 3. Influence of gluon damping on the radiation spectrum

The properties of a medium can significantly influence the probability of bremsstrahlung formation and, thus, the radiation spectrum, see [22] for a comprehensive review. As discussed in [23], the global radiation intensity per unit $x$-interval in the medium changes compared to the reference spectrum in absence of any medium effect (vacuum) in line with the ratio of the corresponding formation lengths, i.e. $dI/dI^0 \simeq l_f/l_f^0$. We adopt a similar scaling law by quantifying the effects of the medium on the radiation spectrum in comparison with the soft $x \ll m_g/m_s$ Gunion-Bertsch (GB) spectrum [24] from a single, high-energy scattering process, i.e. $dI/dI_{GB} \simeq l_f/l_f(x \ll m_g/m_s; \hat{q} \to 0)$. In this way, the influence of multiple, elastic scatterings in the medium as well as the effect of the finite parton mass can be highlighted semi-quantitatively. This is shown in Fig. 2 (left panel): For small $x < m_g/m_s$, multiple scatterings lead to a suppression of the radiation spectrum, i.e. $dI < dI_{GB}$, while with increasing $x$ the influence of the finite parton mass becomes essential. As evident from Fig. 2, the spectrum is stronger suppressed for larger $m_s$.

In an absorptive strongly interacting plasma, bremsstrahlung damp-ing phenomena, as for instance quark–anti-quark pair creation or secondary gluon bremsstrahlung formation, can influence the radiation spectrum as well. These higher-order effects introduce an additional scale in the medium [21, 25]. If this absorption length scale $l_d$ is of the order of or smaller than the formation length of the nascent gluon, as is indicated by the dotted, horizontal lines in the right panel of Fig. 1, then damping phenomena will influence the probability for bremsstrahlung formation in the plasma itself.

In order to highlight this additional effect of an absorptive medium on the radiation spectrum, we exploit as scaling law $dI/dI_{GB} \simeq \tilde{l}_f/l_f(x \ll m_g/m_s; \hat{q} \to 0)$, where $\tilde{l}_f$ denotes the minimum of the two competing length scales $l_f$ and $l_d$. Correspondingly, damping phenomena reduce the radiation spectrum in an intermediate $x$-region when $l_d \leq l_f$. With increasing gluon damping rate $\Gamma \simeq 1/l_d$, the radiation spectrum for a given parton energy becomes more and more affected. This is shown in Fig. 2 (right panel) by employing two different $x$-independent values for $\Gamma$. As visible, the radiation spectrum becomes independent of the parton mass in a large region of $x$. The affected $x$-region becomes also larger with increasing $E$. The reduction of the spectrum is proportional to $1/x$ and, thus, stronger than the suppression due to multiple, elastic scatterings in the medium which is proportional to $1/\sqrt{x}$.

## 4. Qualitative discussion and conclusions

As noted above, damping phenomena drastically influence the radiation spectrum whenever the absorption length scale becomes comparable to or smaller than the gluon formation length, i.e. in particular for large $E$ or large $\Gamma$, both conditions being better fulfilled at LHC, as $\Gamma \propto g^4 T \log(1/g)$ [21]. Nonetheless, a larger quenching parameter $\hat{q}$ significantly reduces $l_f$ and, thus, modifies to some extent the emerging picture, too. In order to understand better where damping effects become important in dependence of the medium parameters and of the parton properties, one may discuss Eq. (2) determining $l_f$ more qualitatively. This can be done, as similarly done in detail in [21], by assuming that either the linear or the quadratic term in $l_f$ contributes dominantly to the condition equation. Defining the gluon formation length as the minimum of the obtained positive solutions gives rise to a fairly good approximation of the exact solution for $l_f$ from Eq. (2).

From a parametric analysis of the competition between the relevant length scales, in analogy to [21], the following picture, as summarized in Fig. 3, emerges in dependence of $\Gamma$ and the Lorentz-factor $\gamma = E/m_s$ in case $m_g^3 > \hat{q}$: For a negligible damping rate, $\Gamma/m_g \to 0$, parts of the differential radiation spectrum are influenced by multiple elastic scatterings in the medium if $\gamma > \gamma_c^{(1)} \sim m_g^3/\hat{q}$. Correspondingly for larger $\hat{q}$, LPM-suppression is relevant in the spectrum already for smaller parton energies. In an absorptive plasma with damping rate $\Gamma < \Gamma_c \sim \hat{q}/m_g^2$, damping effects become important for a part of the differential spectrum in an intermediate $x$-region for $\gamma > \gamma_d^{(1)} \sim \sqrt{\hat{q}/\Gamma^3}$. This is shown by region I in Fig. 3. With increasing $\Gamma$, this part becomes larger and for $\Gamma \to \Gamma_c$ one finds $\gamma_d^{(1)} \to \gamma_c^{(1)}$ such that the region in which damping effects are dominant eventually overlaps totally with the region originally affected by LPM-suppression. For even larger $\Gamma > \Gamma_c$, damping phenomena can become influential already for $\gamma > \gamma_d^{(2)} \sim m_g/\Gamma > 1$, where $\gamma_d^{(2)} < \gamma_c^{(1)}$, as shown by region II in Fig. 3. This shows impressively that damping effects have an impact on the radiation spectrum off a parton with mass $m_s$ for large $\Gamma$ and/or large $E$.

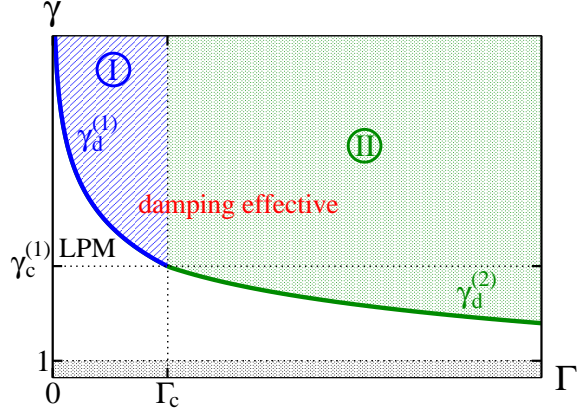

Fig. 3. Illustration of the regions in $\Gamma$-$\gamma$-space, in which damping phenomena become important for parts of the radiation spectrum (regions I and II), see text for details. Also shown is the region $\gamma_c^{(1)} < \gamma < \gamma_d^{(1)}$ for $\Gamma < \Gamma_c$ (region LPM), in which LPM-suppression is of relevance. While for $\Gamma \to 0$, $\gamma_d^{(1)} \to \infty$, one finds $\gamma_d^{(1)} \to \gamma_c^{(1)}$ for $\Gamma \to \Gamma_c$ such that with increasing $\Gamma$ the usual (undamped) LPM-regime shrinks and eventually disappears.

In summary, we have shown that gluon damping in a hot quark-gluon plasma could affect significantly the established radiation pattern for HQ (see [26] for a review). Consequences on HQ observables in URHIC will be addressed in a future contribution. Let us insist that the picture presented here for medium-induced gluon radiation only applies when the formation length $l_f$ is smaller than the path length $L$; otherwise, interference with the initial bremsstrahlung should be included in the analysis. However, we conjecture that gluon damping is still a major effect if $1/\Gamma \lesssim L \lesssim l_f$.